\documentclass[a4paper,accepted=2024-11-20]{quantumarticle}
\pdfoutput=1
\usepackage{mathtools}        
\usepackage{bm}
\usepackage{bbm}
\usepackage[square,numbers,sort&compress]{natbib}
\usepackage{float}            
\usepackage{graphicx}         
\usepackage[dvipsnames, table]{xcolor}
\usepackage[caption=false]{subfig}
\usepackage{physics}
\usepackage{relsize}          
\usepackage{epstopdf}
\usepackage{soul}
\usepackage{hyperref}
\hypersetup{
	linkcolor=blue,
	citecolor=blue,
	filecolor=blue,
	urlcolor=blue,
	colorlinks=true
}

\usepackage{lipsum}

\newcommand{\UIUCPHYS}[0]{Department of Physics, University of Illinois at Urbana-Champaign, Urbana, IL 61801, USA}
\newcommand{\USRA}[0]{USRA Research Institute for Advanced Computer Science, Mountain View, California 94043, USA}
\newcommand{\NASA}[0]{Intelligent Systems Division, NASA Ames Research Center, Moffett Field, California 94035, USA}
\newcommand{\KBR}[0]{KBR, Inc., NASA Ames Research Center, Moffett Field, California 94035, USA}

\allowdisplaybreaks 

\begin{document}

\title{Beyond MP2 initialization for unitary coupled cluster quantum circuits}

\author{Mark R. Hirsbrunner}\email{hrsbrnn2@illinois.edu}\affiliation{\UIUCPHYS}\affiliation{\USRA}
\author{Diana Chamaki}\affiliation{\USRA}
\author{J. Wayne Mullinax}\affiliation{\KBR}\affiliation{\NASA}
\author{Norm M. Tubman}\email{norman.m.tubman@nasa.gov}\affiliation{\NASA}

\begin{abstract}
    The unitary coupled cluster (UCC) ansatz is a promising tool for achieving high-precision results using the variational quantum eigensolver (VQE) algorithm in the NISQ era. However, results on quantum hardware are thus far very limited and simulations have only accessed small system sizes. We advance the state of the art of UCC simulations by utilizing an efficient sparse wavefunction circuit solver and studying systems up to 64 qubits. Here we report results obtained using this solver that demonstrate the power of the UCC ansatz and address pressing questions about optimal initial parameterizations and circuit construction, among others. Our approach enables meaningful benchmarking of the UCC ansatz, a crucial step in assessing the utility of VQE for achieving quantum advantage.
\end{abstract}

\maketitle

\textit{Introduction}.--- Simulating many-body fermionic systems is a promising future application of quantum computing~\cite{Feynman1982, Abrams1997, Ortiz2001}. While it is not yet clear that quantum advantage can generically be achieved in this area~\cite{Lee2022}, it is believed that phase estimation can solve ground state problems for molecular systems that are beyond the reach of classical computers. However, it remains an open question whether or not other approaches can achieve quantum advantage with fewer resources~\cite{Kitaev1995, Abrams1999, Aspuru-Guzik2005, McArdle2020}. Phase estimation and other algorithms benefit from, or even require, significant overlap between the trial quantum state and the true solution. Single Slater determinants, such as Hartree-Fock states~\cite{szabo2012modern}, are often used as the trial state when solving for ground states, as they are assumed to produce a sufficiently large overlap with the ground state wavefunction in many cases~\cite{Tubman2018,O_Brien2019,O_Malley2016}. Yet such single determinant states may not be sufficient for arbitrarily large system sizes~\cite{Vleck1936, McClean2014, Tubman2018}. Improving quantum state preparation techniques is a key step toward advancing quantum computing for quantum chemistry and other Hamiltonian simulation applications.

\begin{figure*}[t]
    \subfloat[]{\label{fig:1a}\includegraphics[width=.49\linewidth]{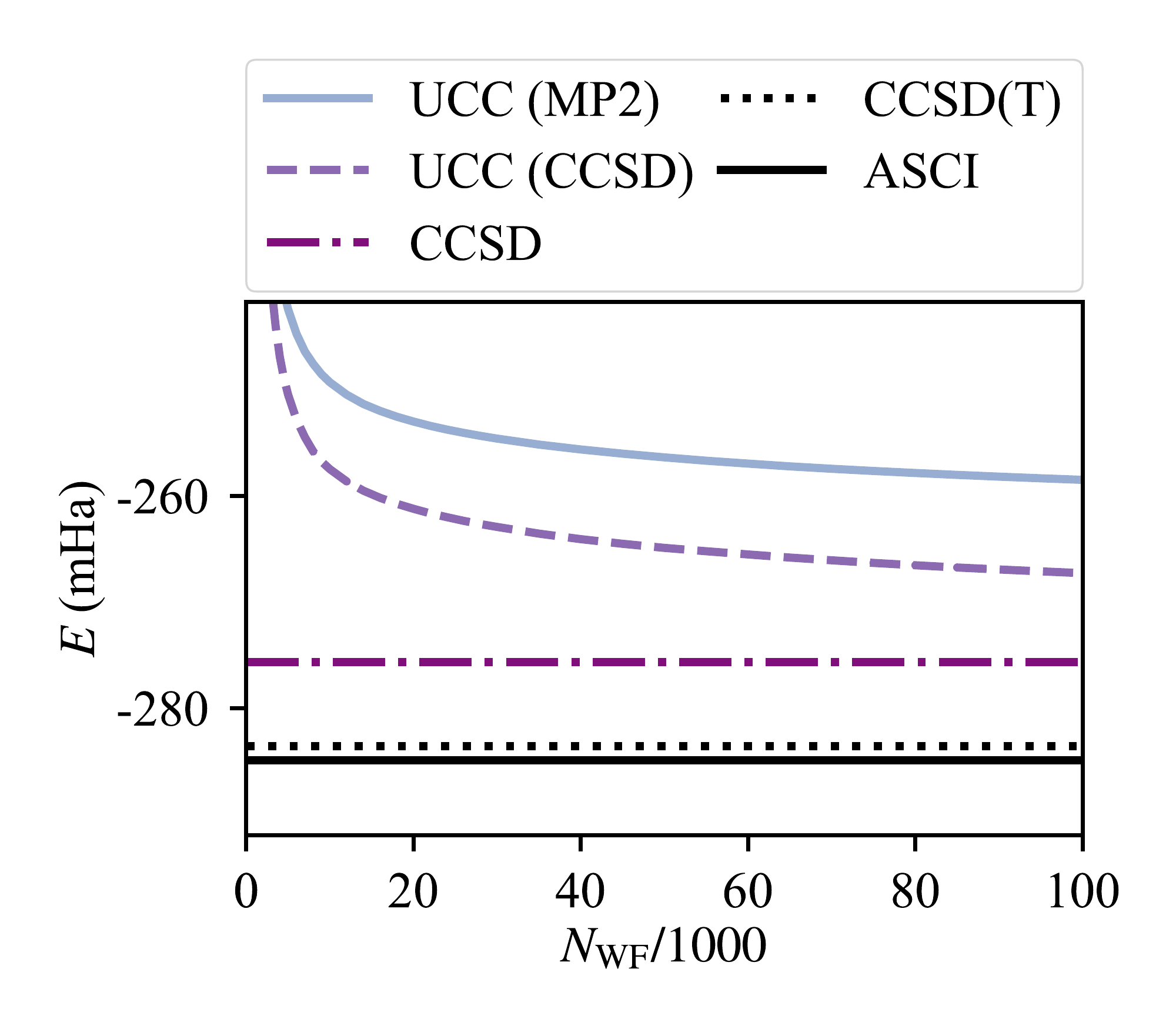}}
    \hfill
    \subfloat[]{\label{fig:1b}\includegraphics[width=.49\linewidth]{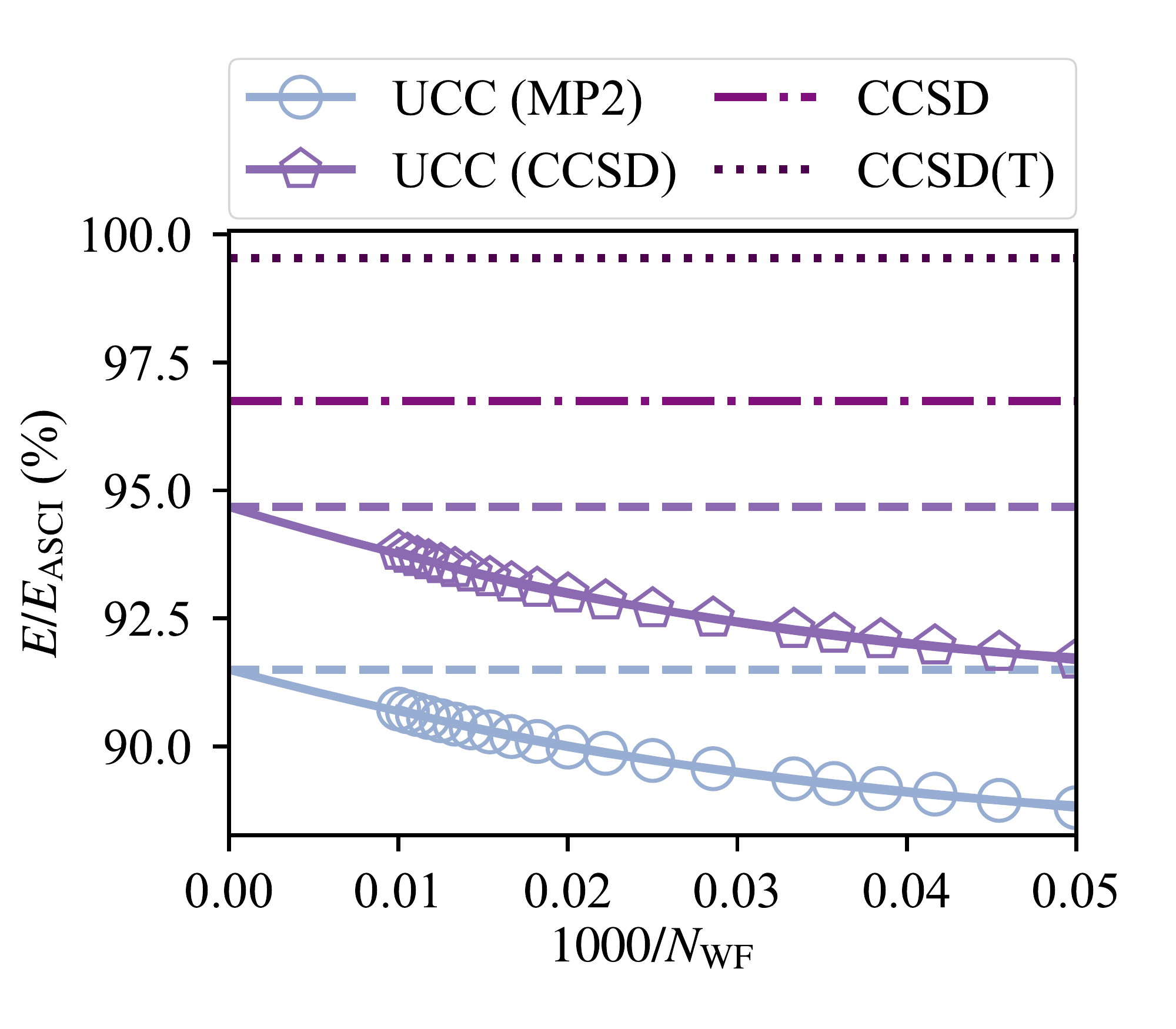}}
    \hfill
    \caption[]{
    (a) The UCC (MP2) (light blue solid line) and UCC (CCSD) (light purple dashed line) correlation energies of $\mathrm{CH}_2\mathrm{O}$ as functions of $N_{\mathrm{WF}}$. The dark purple dot-dashed and black dotted lines denote the CCSD and CCSD(T) correlation energies, respectively, and the black solid line marks the ASCI energy calculated using $10^6$ determinants. 
    (b) The UCC (MP2) (circles) and UCC (CCSD) (pentagons) correlation energies as a fraction of the CCSD(T) correlation energy plotted versus $1/N_{\mathrm{WF}}$. The solid lines are quadratic fits to the UCC (CCSD) and UCC (MP2) data, fitted using the twenty data points with the largest $N_{\mathrm{WF}}$. The dashed lines mark the $y$-intercepts of the fits. The dot-dashed and dotted lines indicate the ratio of the CCSD and CCSD(T) correlation energies to the ASCI correlation energy.
    }
    \label{fig:1}
\end{figure*}

Noise and decoherence present another central difficulty of achieving quantum advantage in the current noisy intermediate-scale quantum (NISQ) era  of quantum hardware~\cite{Preskill2018}. 
The variational quantum eigensolver (VQE) is a quantum-classical hybrid algorithm that is particularly well-suited for NISQ devices and has a wide range of applicability~\cite{ Stavenger2022,Bassman2022,Magann2023,Aram2021,Anastasiou2022,Carrasquilla2021,Martyn2022,Burton2022,Claudino2020,Tang2021,Chamaki2022,Ayral2022,Cerezo2021,Luo2022,Guerreschi2020,Smelyanskiy2016,Cao2019,Smith2022,Lu2022,Zhao2020,Pathak2022,Kandala2017,Chamaki2022-1,Lin2020,Unmuth2022,Sherbert2021,Sherbert2022,Traps2022,Oganov2020,Miro2020,Kirby2021,Kirby2019,Li2021,Jahin2022, Huggins2020}. While VQE does not provide exact ground state solutions like quantum phase estimation, the approximate wavefunctions produced by VQE are often sufficiently accurate to provide meaningful physical insights. Furthermore, these approximate solutions are well-suited for quantum state preparation for use in more accurate algorithms.

Despite its current popularity, VQE possesses a number of drawbacks. In particular, the classical optimization of circuit parameters presents many challenges, including barren plateaus (i.e., exponentially vanishing gradients in high dimensions), local minima, and saddle points~\cite{Huembeli2021, McClean2018,Bittel2021,Uvarov2020, Arrasmith2021}. Many approaches exist for minimizing the computational burden of classical optimization for VQE, with some proposals eschewing optimization entirely~\cite{Kremenetski2021-1,vlad2021,Huggins2020, Baek2022, ravi2022cafqa}. The crux of several of these strategies is a focus on choosing high-quality initial parameters, shifting some of the computational burden from optimization to initialization. While it seems unlikely with current technology for classical computation to produce a circuit capable of generating wave functions beyond what we can simulate classically, approaching this limit with high performance computing resources is likely to be a fruitful strategy for the near term and for the fault tolerant era. In this work we compare the utility of different initialization strategies for a particular VQE ansatz that is often employed in quantum chemistry problems, the unitary coupled cluster (UCC) ansatz~\cite{Hoffmann1988, Bartlett1989,Peruzzo2014,Yung2014, Romero2019, Anand2022}.

There are several proposed strategies for generating the initial parameters for the UCC ansatz~\cite{Moller1934, McClean2016, Romero2019,Tilly2021,Fed2021}, including applications in which no optimization is performed on quantum hardware~\cite{Baek2022}.  The most widely employed strategy generates parameters using classical Møller–Plesset perturbation theory of the second order (MP2)~\cite{Moller1934, McClean2016, Romero2019,Tilly2021}.  Another less thoroughly studied approach is the use of the coupled cluster singles and doubles (CCSD) classical simulation method to generate initial parameters~\cite{Coester1960,Bartlett2007}. The CCSD technique generally produces more accurate ground state energies than MP2 calculations and is not prohibitively more computationally burdensome for all but the largest of problems. However, CCSD has not been previously explored as a method to initialize VQE circuits, partly because the sparse wave function simulators in which it can be tested were only recently developed, and partly because it often produces non-physical energies for strongly correlated molecules. This raises the question: Which technique produces superior initial parameters for UCC ansatzes, MP2 or CCSD?

In this paper, we provide the first numerical study (to our knowledge) comparing the performance of UCC ansatzes prepared using parameters generated via MP2 and via CCSD. We employ an algorithm for the factorized form of UCC implemented using our state-of-the-art sparse wavefunction circuit solver, enabling us to study problems of up to 64 qubits~\cite{Chen2020, mullinax2022}. By calculating the ground state energy of a wide range of molecules using both MP2 and CCSD parameters in the UCC ansatz, we show conclusively that CCSD parameters outperform MP2, generating significantly more accurate ground state energies. Importantly, we find that CCSD parameters outperform MP2 even in the case that the CCSD calculation fails to produce a physical energy. We also compare the subsequent optimization of circuits initialized with MP2 and CCSD parameters, finding that CCSD circuits generally requires far fewer optimization steps to obtain the same accuracy as MP2 circuits.
    
\textit{Technique}.--- The UCC ansatz is an exponential operator acting on the Hartree--Fock reference wavefunction defined as 
\begin{equation}
    \label{eq:ucc}
    \ket{\Psi_{\mathrm{UCC}}} = \exp( \hat{T} - \hat{T}^{\dagger})\ket{\Psi_{0}}
\end{equation}
where the coupled cluster operator $\hat{T}$ is 
\begin{equation}
    \label{eq:t}
    \hat{T} = \sum_{i}^{\mathrm{occ}} \sum_{a}^{\mathrm{vir}} \theta_{i}^{a}\hat{a}_{a}^{\dagger}\hat{a}_{i}
            + \sum_{ij}^{\mathrm{occ}} \sum_{ab}^{\mathrm{vir}} \theta_{ij}^{ab}\hat{a}_{a}^{\dagger} \hat{a}_{b}^{\dagger} \hat{a}_{j} \hat{a}_{i} + \cdots
\end{equation}
The $\hat{a}^{\dagger}$ and $\hat{a}$ operators are the second-quantized creation and annihilation operators, respectively, acting on the occupied molecular orbitals in the reference wavefunction indexed by $i,j,\ldots$ or the virtual orbitals indexed by $a,b,\ldots$. The parameters of the UCC ansatz are indicated by $\theta$. We employ the factorized form of the UCC ansatz, which is given by
\begin{equation}
    \label{eq:ansatz}
    \ket{\Psi_{\mathrm{UCC}}} = \prod_{ij\cdots}^{\mathrm{occ}} \prod_{ab\cdots}^{\mathrm{vir}} \hat{U}^{ab\cdots}_{ij\cdots}\ket{\Psi_{0}},
\end{equation}
where the individual UCC exponential factors are defined as
\begin{equation}
    \label{eq:factor}
    \hat{U}^{ab\cdots}_{ij\cdots} = \exp(\theta_{ij\cdots}^{ab\cdots}(\hat{a}_{ij\cdots}^{ab\cdots} - \hat{a}_{ab\cdots}^{ij\cdots})).
\end{equation}
We only include single excitations ($\hat{a}^{a}_{i} = \hat{a}^{\dagger}_a\hat{a}_i$) and double excitations ($\hat{a}^{ab}_{ij} = \hat{a}^{\dagger}_{a}\hat{a}^{\dagger}_{b}\hat{a}_{j}\hat{a}_{i}$) in the ansatz, along with the conjugate deexcitation operators $\hat{a}^{i}_{a}$ and $\hat{a}^{ij}_{ab}$. This approximation to the full UCC ansatz is known as the unitary coupled cluster singles and doubles (UCCSD) ansatz. Different qubit to fermion mappings can be used to realize these operators in circuit form, but they do not affect our results or change the ansatz expressivity.

We utilize a specific representation of the UCCSD ansatz that exploits the fact that each UCC factor $\hat{U}^{ab\cdots}_{ij\cdots}$ can be expressed in terms of sines and cosines of the parameters that can be efficiently computed on classical hardware~\cite{Chen2020}. The order of the individual UCCSD factors is not strictly defined~\cite{grim2019, Grimsley:2019:3007}, and we chose to order them based on the magnitude of the parameter values ($\abs{\theta}$), placing the factor $\hat{U}_{ij\cdots}^{ab\cdots}$ that contains the largest parameter to the right in Equation \eqref{eq:ansatz}. We refer to this as the ``magnitude'' ordering~\footnote{The parameters produced by MP2 and CCSD are between $-\pi$ and $\pi$ for all molecules we study, so the periodicity of $\theta$ does not cause any ambiguity in the ``magnitude'' ordering.}.

We generate the conventional MP2 and CCSD UCCSD parameters using the PySCF implementation of the techniques described in~\cite{McClean2016}, and note that the MP2 parameterization does not include any single excitation operators~\cite{Sun2020}. We use a computationally efficient sparse wavefunction approach, limiting the number of determinants included in the wavefunction after each UCC factor is applied~\cite{mullinax2022}. We do this by checking the number of determinants $N$ in the wavefunction after applying each UCC factor. If $N$ is greater than the desired number of determinants, $N_{\textrm{WF}}$, we sort the amplitudes by magnitude and discard the determinants with the smallest amplitudes such that only $N_{\textrm{WF}}$ determinants remain in the wavefunction.

\begin{figure}
    \centering
    \includegraphics[width=\linewidth]{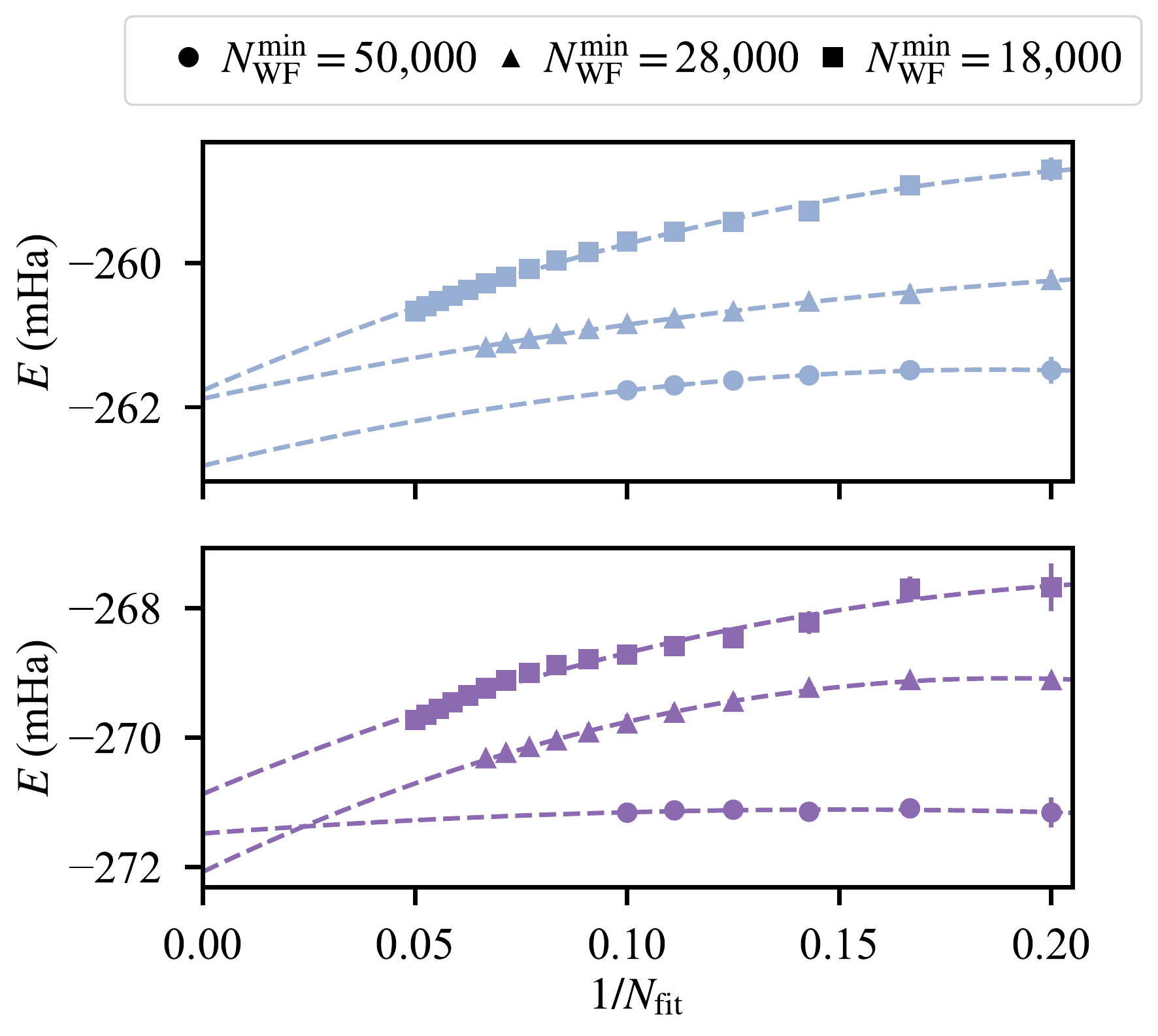}
    \caption[]{The (top) UCC (MP2) and (bottom) UCC (CCSD) correlation energies of $\mathrm{CH}_2\mathrm{O}$ obtained by our extrapolation in Eq.~\eqref{eq:extrap} for different choices of fitting windows. Each data point the energy extrapolated using a fitting window consisting of the $N_{\mathrm{fit}}$ smallest values of $N_{\mathrm{WF}}$ that are greater than a particular minimum value, $N_{\mathrm{WF}}^{\mathrm{min}}$. The error bars (which are too small to be visible for most data points) correspond to the standard deviation obtained from the extrapolation procedure. The dashed lines are quadratic fits of the extrapolated energies as functions of $1/N_{\mathrm{fit}}$.}
    \label{fig:fit_analysis}
\end{figure}

\textit{Results}.--- We report the correlation energies obtained from UCCSD circuits parameterized using MP2 and CCSD parameters for a wide range of molecules.  All orbitals used for both the classical and circuit calulcations come from Hartree-Fock simulations.   For the molecules LiH, HF, $\mathrm{NH}_3$, $\mathrm{CH}_4$, $\mathrm{H}_2\mathrm{O}$, $\mathrm{N}_2$, $\mathrm{F}_2$, and $\mathrm{CH}_2\mathrm{O}$ we use experimental geometries from the CCCBDB database and employ the cc-pCVDZ basis set~\cite{CCCBDB}. We also study the linear hydrogen chains $\mathrm{H}_8$, $\mathrm{H}_{10}$, $\mathrm{H}_{12}$, and $\mathrm{H}_{14}$, for which we use an interatomic distance of 1~\AA, and a stretched geometry of $\mathrm{H}_{10}$ with an interatomic distance of 1.5~\AA. We employ the STO-6G basis set for all hydrogen chains. Our sparse wavefunction circuit solver is limited to 64 qubits, so we include only the 32 lowest-energy molecular orbitals in each molecule~\footnote{This limitation of our solver is not algorithmic and future implementations can be expanded beyond 64 qubits.}. We also limit the total number of doubles operators in the UCC circuits for these molecules to 15,000. This significantly reduces the computational cost of the calculations but still greatly exceeds the number of operators required to converge the energy produced by the circuits. Our approach has similar scaling to a time-dependent selected configuration interaction approach, which some of us have applied to larger systems in other contexts~\cite{Kremenetski2021-1,vlad2021}.

Because we limit the number of determinants retained in the wavefunction to $N_{\mathrm{WF}}$, we must study the dependence of the correlation energies on $N_{\mathrm{WF}}$ and extrapolate to the large-$N_{\mathrm{WF}}$ limit to obtain a converged result. Specifically, we calculate the correlation energy as a function of $N_{\mathrm{WF}}$ up to $N_{\mathrm{WF}}=$100,000 for each molecule, as shown in Fig.~\ref{fig:1a} for $\mathrm{CH}_2\mathrm{O}$. We extrapolate to the large-$N_{\mathrm{WF}}$ limit via a quadratic fitting of the data as a function of $N_{\mathrm{WF}}^{-1}$,
\begin{equation}
E = a N_{\mathrm{WF}}^{-2} + bN_{\mathrm{WF}}^{-1} + c,
\label{eq:extrap}
\end{equation}
as shown in Fig.~\ref{fig:1b}. The $y$-intercept of the quadratic fit is the extrapolated correlation energy that would be obtained if we pruned no determinants during the calculation. Thus this is a prediction of the energy that would be produced on perfect quantum hardware. We refer to these extrapolated energies as the UCC (MP2) and UCC (CCSD) correlation energies, depending on the initial parameters used in the circuit. The fit accounts for the twenty data points at the largest values of $N_{\mathrm{WF}}$, those with $N_{\mathrm{WF}}$ greater than 18,000, which produces extrapolated energies with insignificant errors of at most 0.1~mHar.

The choice of fitting only the twenty largest values of $N_{\mathrm{WF}}$ is arbitrary, so in Fig.~\ref{fig:fit_analysis} we study the dependence of the extrapolated energy of $\mathrm{CH}_2\mathrm{O}$ on the choice of the fitting window. We calculate the extrapolated energy using fitting windows containing a range of $N_{\mathrm{fit}}$ smallest values of $N_{\mathrm{WF}}$ that are greater than a particular minimum value, $N_{\mathrm{WF}}^{\mathrm{min}}$. For each value of $N_{\mathrm{WF}}^{\mathrm{min}}$, we perform a secondary extrapolation of the UCC (MP2) and UCC (CCSD) energies to the infinite-$N_{\mathrm{fit}}$ limit, marked by the dashed lines. Both the secondary extrapolated energies and the values obtained at the maximum value of $N_{\mathrm{fit}}$ vary by less than 1.6 mHa between the different values of $N_{\mathrm{WF}}^{\mathrm{min}}$, indicating that varying the energy window produces extrapolated energies that are consistent within chemical accuracy. Further study is required to confirm that this holds in general and to determine the best practice for extrapolating energies to the infinite-$N_{\mathrm{WF}}$ limit. However, we find the conservative energies predicted by our chosen fitting window to be sufficiently accurate for the context of this work.

\begin{table*}
\addtolength{\tabcolsep}{-0.0pt}
    \centering
    \caption{The FCI, CCSD(T), CCSD, UCC (CCSD), MP2, and UCC (MP2) correlation energies of the hydrogen chains and LiH. The UCC energies are obtained via the fitting procedure shown in Fig.~\ref{fig:1}. The row labeled $\mathrm{H}_{10}^*$ uses a stretched geometry with an interatomic distance of 1.5~\AA. All energies are reported as absolute values and in units of milliHartrees.}
    {
    \renewcommand{\arraystretch}{1.2}
\begin{tabular}{ccccccc}
 Mol & FCI & CCSD(T) & CCSD    & UCC (CCSD)  & MP2  & UCC (MP2)      \\
\hline
$\mathrm{H}_8$ &  134.68  & 134.65    &   133.60     &   133.00    & 85.19 & 111.69   \\
$\mathrm{H}_{10}$ &  167.78  & 167.64    &   165.77     &   164.86   & 107.62   & 139.08   \\
$\mathrm{H}_{10}^*$ &  403.81  & 434.55    &   426.50     &   354.74   & 208.45   & 293.67  \\
$\mathrm{H}_{12}$ &  200.90  & 200.62    &   197.81     &   196.62   & 130.27   & 166.44    \\
$\mathrm{H}_{14}$ &  234.05  & 233.60    &   229.75     &   228.92   & 153.11   & 194.45  \\
LiH &  64.75  & 64.74    &   64.69     &   64.69   & 51.80  & 61.28   \\
\\
\end{tabular}
    }
    \label{table:1}
\end{table*}

\begin{figure}
    \includegraphics[width=0.95\linewidth]{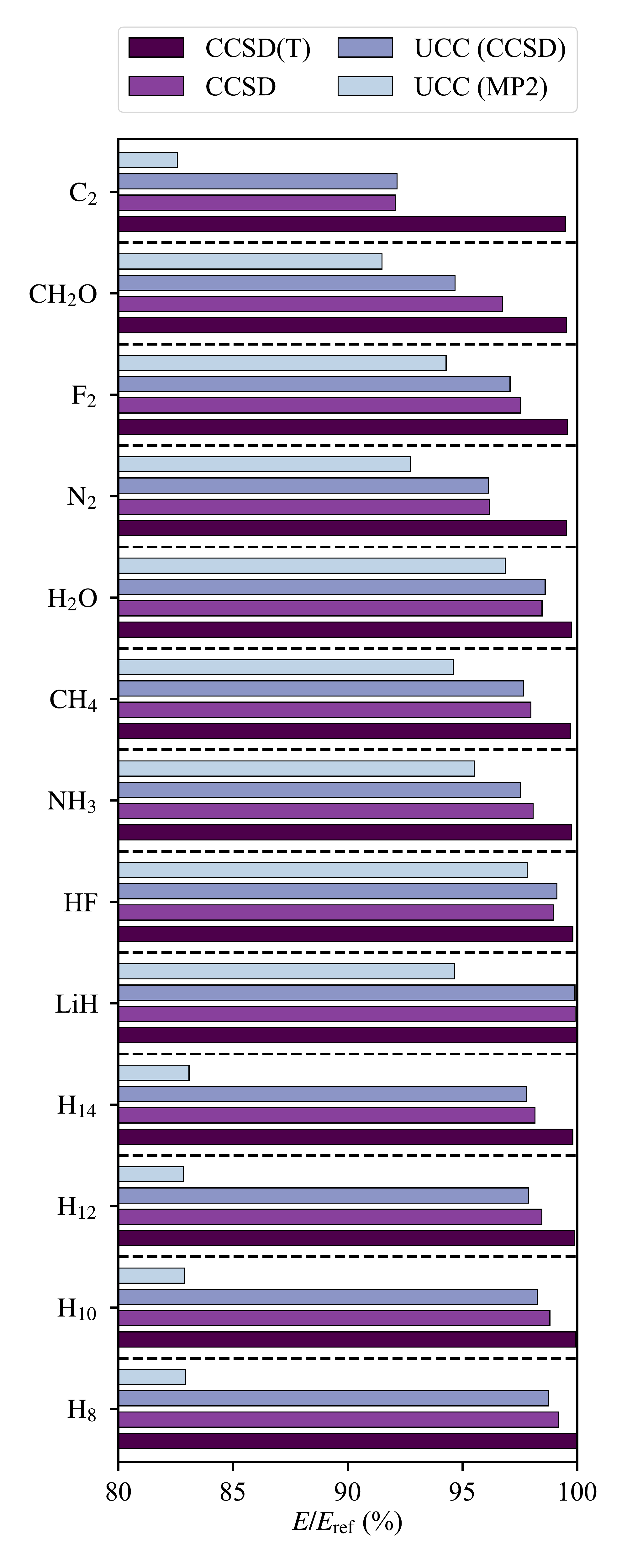}
    \caption[]{The CCSD(T), CCSD, UCC (CCSD), and UCC (MP2) correlation energies as percentages of the best available reference energy (FCI for hydrogen chains and LiH, ASCI otherwise).}
    \label{fig:2}
\end{figure}

We report the CCSD(T), CCSD, UCC (MP2), UCC (CCSD), and full configuration interaction (FCI) correlation energies for the hydrogen chains and LiH in Table~\ref{table:1}. Calculating the FCI energy for the remaining molecules is impractical, so we instead report the adaptive sampling configuration interaction (ASCI) correlation energies~\footnote{We use $10^6$ determinants for all ASCI calculations, at which the energies are converged.} for these molecules in Table~\ref{table:2}, along with the CCSD(T), CCSD, UCC (MP2), and UCC (CCSD) correlation energies~\cite{tubman2020}. We also plot these energies as fractions of the best reference energy, either FCI or ASCI, for each molecule in Fig.~\ref{fig:2}. The UCC (CCSD) energy outperforms the UCC (MP2) energy by a wide margin in all cases, with a difference of approximately 15\% of the reference energy for the hydrogen chains (including stretched $\mathrm{H}_{10}$) and differences ranging from 1.3\% to 9.6\% for the remaining molecules.

Because the individual terms in the factorized form of the UCCSD ansatz do not necessarily commute, the ordering of the operators can have a significant impact on the accuracy of the ansatz~\cite{grim2019}. To address this concern, we calculate the correlation energy of the molecules we study in this work using 100 random orderings of the UCCSD factors. We find that the standard deviation of the correlation energy is at most 0.1 mHa for molecules with equilibrium geometries, with the exception of $\mathrm{C}_2$ which produced a standard deviation of 0.4 mHa. Only the strongly correlated stretched geometry of $\mathrm{H}_{10}$ has a significant standard deviation of 2.4 mHa. We set $N_{\mathrm{WF}}$ to 10,000 for these calculations to reduce the computational burden, which likely artificially inflates the standard deviations. We conclude that factor ordering is significant only for strongly correlated molecules, in agreement with previous studies~\cite{grim2019}. Importantly, we find that the magnitude ordering obtains energies close to the minimum energy produced by random orderings for all molecules besides $\mathrm{CH}_2\mathrm{O}$, for which the magnitude ordering produced an energy approximately 0.15 mHa above the minimum.

\begin{figure}
    \includegraphics[width=0.95\linewidth]{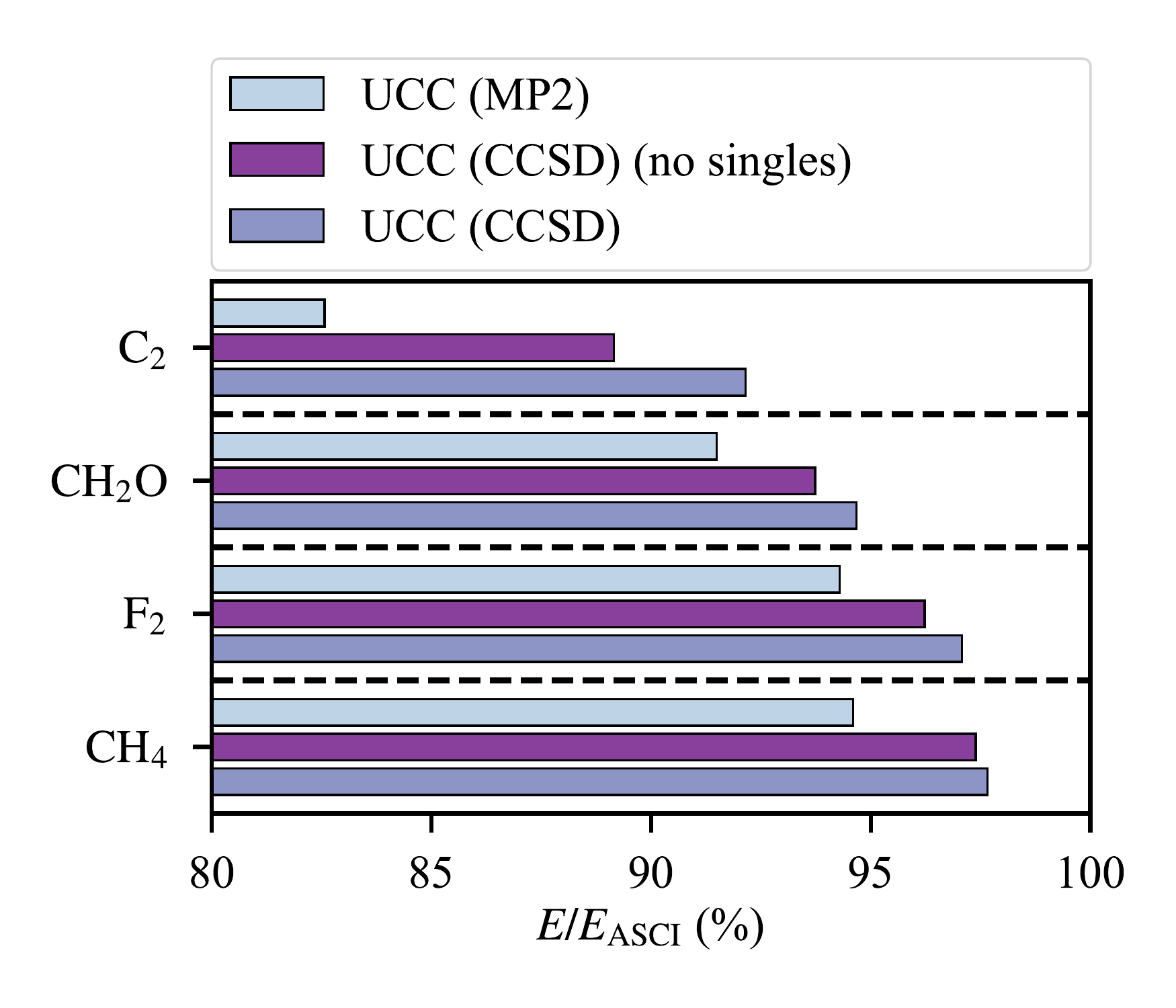}
    \caption[]{A comparison of the extrapolated energies for $\mathrm{CH}_4$, $\mathrm{F}_2$, $\mathrm{CH}_2\mathrm{O}$, and $\mathrm{C}_2$ obtained using UCC circuits initialized with MP2 parameters, CCSD parameters, and CCSD parameters excluding singles operators, all presented as percentages of the ASCI energy. While removing the singles operators lowers the accuracy of the UCC (CCSD) energies, they still outperform the UCC (MP2) energies by a significant margin.}
    \label{fig:no_singles}
\end{figure}

The UCC (CCSD) energy closely matches the CCSD energy for all molecules studied, with the exception of stretched $\mathrm{H}_{10}$, but the MP2 and UCC (MP2) energies do not exhibit any such close agreement. Excluding the results for stretched $H_{10}$, the differences between the CCSD and UCC (CCSD) energies range between 0.0\% and 2.1\% with an average of 0.4\% for the molecules we study, while for MP2 and UCC (MP2) they range between 0.4\% and 19.7\% with an average of 9.2\%. These statistics show a clear advantage for UCC (CCSD). Furthermore, the UCC (CCSD) energies for HF, $\mathrm{H}_2\mathrm{O}$, and $\mathrm{C}_2$ are \emph{lower} than the corresponding CCSD energies. As such, the CCSD parameterization is likely better suited than MP2 for use in recent proposals for no-optimization strategies to obtain quantum advantage~\cite{Baek2022}. It is plausible that the better performance of the CCSD initialization could arise solely from the fact that the MP2 parameterization does not include any singles operators. To determine whether or not this is the case, we repeated our analysis for a subset of the molecules studied above, now excluding the singles operators from the CCSD-initialized UCC circuits, and plot the results in Fig.~\ref{fig:no_singles}. While the resulting extrapolated energies do not achieve the same level of accuracy as standard UCC (CCSD), they still outperform the MP2-initialized circuits, indicating that the advantage is not simply due to the inclusion of singles operators.

\begin{table*}
\addtolength{\tabcolsep}{-0.0pt}
    \centering
    \caption{The ASCI, CCSD(T), CCSD, UCC (CCSD), MP2, and UCC (MP2) correlation energies of the larger molecules for which FCI is impractical. The UCC energies are obtained via the fitting procedure shown in Fig.~\ref{fig:1}. All energies are reported as absolute values and in units of milliHartrees.}
    {
    \renewcommand{\arraystretch}{1.2}

\begin{tabular}{ccccccc}
 Mol & ASCI & CCSD(T) & CCSD    & UCC (CCSD) & MP2 & UCC (MP2)      \\
\hline
HF & 251.24 & 250.75  & 248.61    &   249.03     &   242.84 & 245.76          \\
$\mathrm{NH}_3$ & 239.02 & 238.42  & 234.42    &   233.12     & 217.13 &   228.29     \\
$\mathrm{CH}_4$ & 183.28 & 182.72  & 179.58    &   178.98   & 156.41  &   173.38        \\
$\mathrm{H}_2\mathrm{O}$ & 255.71 & 255.09  & 251.79    &   252.14   & 241.37  &   247.68        \\
$\mathrm{N}_2$ & 365.24 & 363.53  & 351.26    &   351.13     & 347.65 &   338.73          \\
$\mathrm{F}_2$ & 456.66 & 454.75  & 445.42    &   443.28     &   436.34 & 430.58          \\
$\mathrm{CH}_2\mathrm{O}$ & 284.90 & 283.59  & 275.63    &   269.73  & 261.75   &   260.67          \\
$\mathrm{C}_2$ & 382.23 & 380.23  & 351.91    &   352.21   & 350.27  &   315.59       \\
\\
\end{tabular}

    }
    \label{table:2}
\end{table*}

It is well-known that traditional (projected) classical coupled cluster techniques can, in some situations, obtain energies that are not variational (dropping below the FCI results) or, even worse, energies that diverge from the physical ground state result. One such failure scenario is the chemistry of bond breaking, which we investigate here using the $\mathrm{H}_{10}$ molecule with a stretched interatomic distance of 1.5~\AA. The CCSD and CCSD(T) energies of this molecule are lower than the FCI energy, an example of coupled cluster techniques producing non-variational results. Despite the failure of CCSD to produce an accurate energy for this molecule, the UCC circuit parameterized with CCSD produces a valid energy, as it must because the VQE approach is variational. The UCC (CCSD) energy for stretched $\mathrm{H}_{10}$ is 12.2\% higher than the FCI energy, compared to 1.7\% higher for the equilibrium geometry. These results show that the UCC ansatz parameterized with CCSD is robust to failures of the classical theory, but with some loss of accuracy. Regardless, our results show a close correspondence between UCC (CCSD) and CCSD theories and further study of this will provide insight into the power of coupled cluster approaches on quantum hardware.

\begin{figure*}
    \centering
    \subfloat[][LiH]{\includegraphics[width=0.3\linewidth]{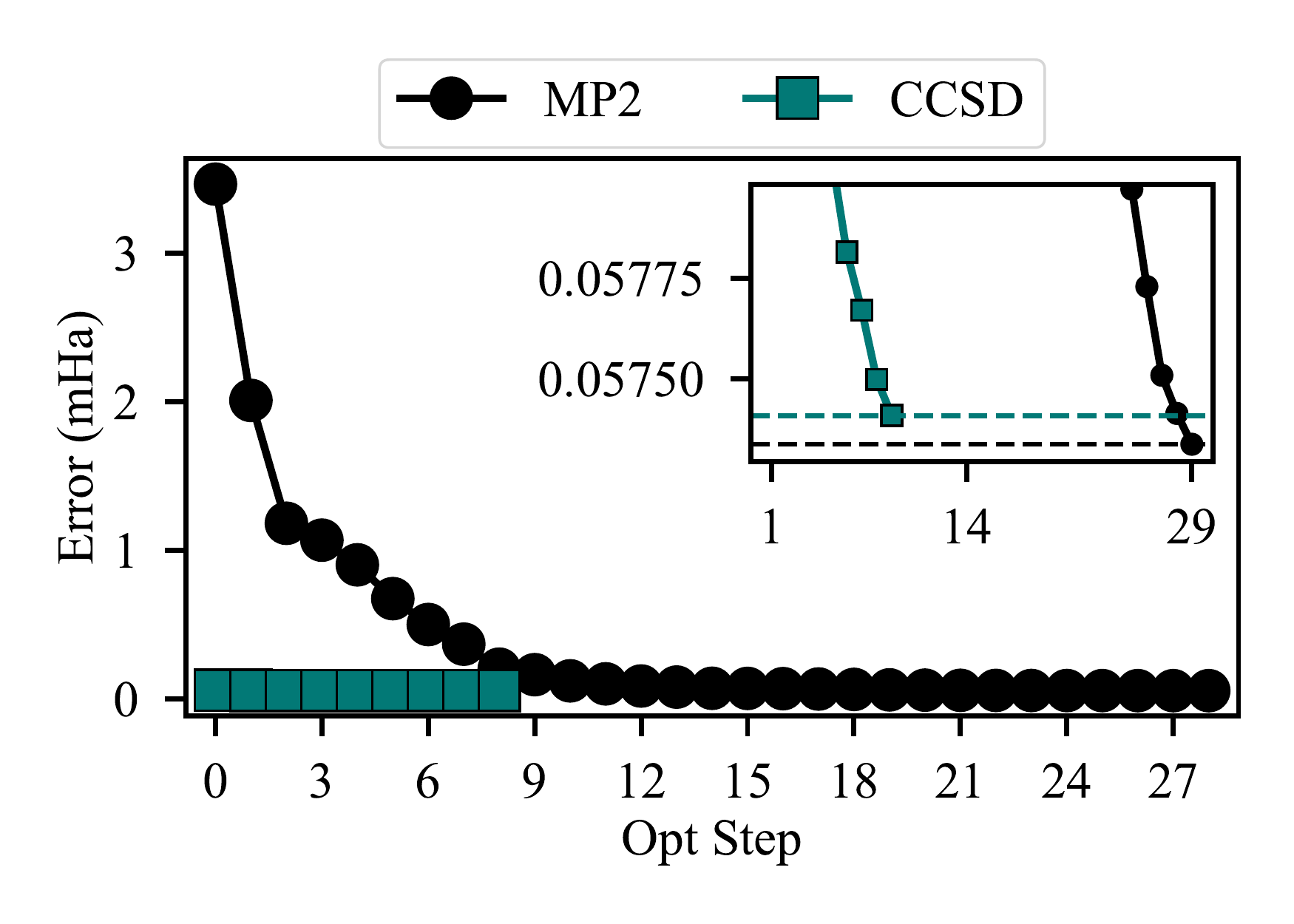}}
    \subfloat[][$\mathrm{BeH}_2$]{\includegraphics[width=0.3\linewidth]{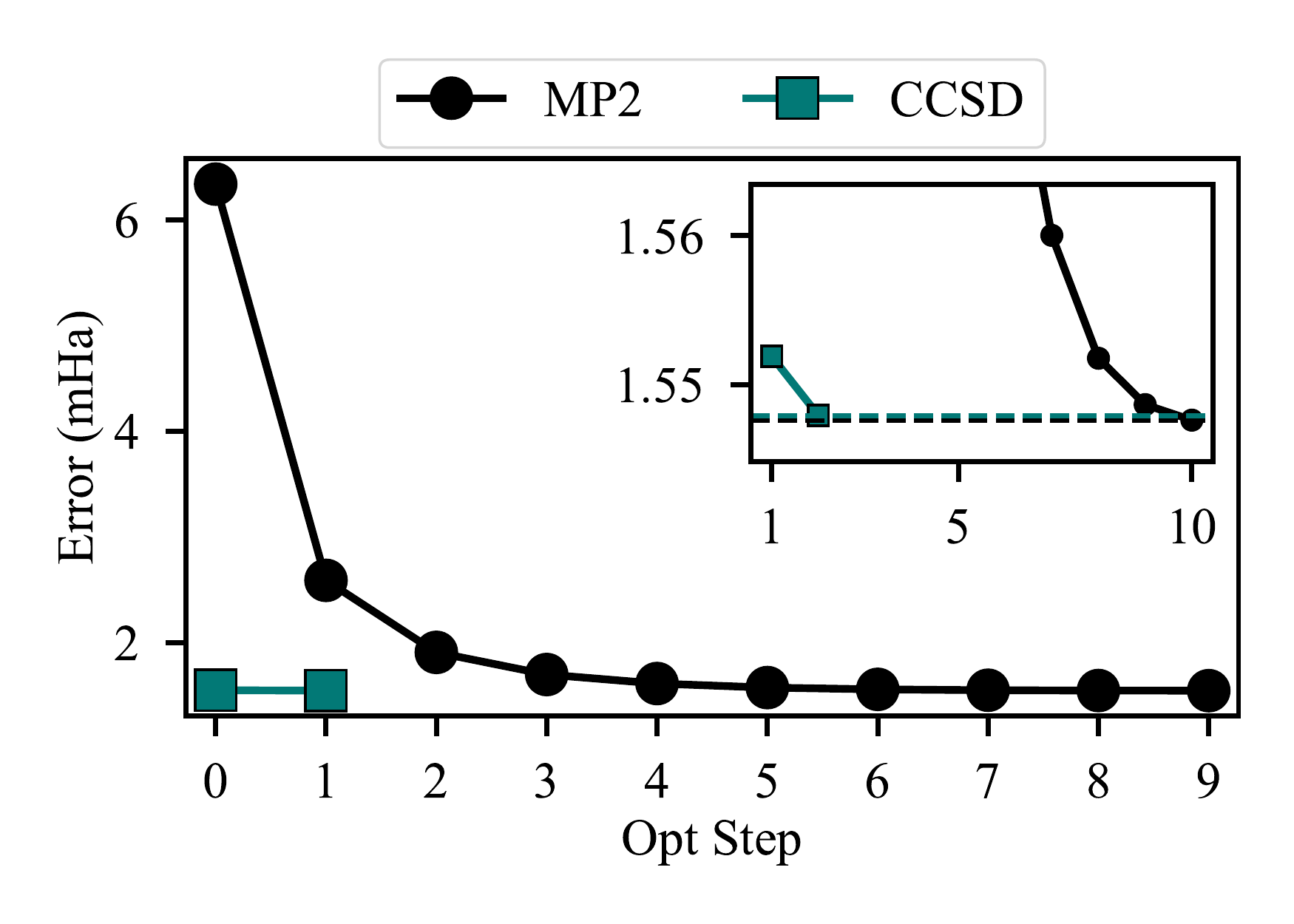}}
    \subfloat[][Stretched $\mathrm{H}_{10}$]{\includegraphics[width=0.3\linewidth]{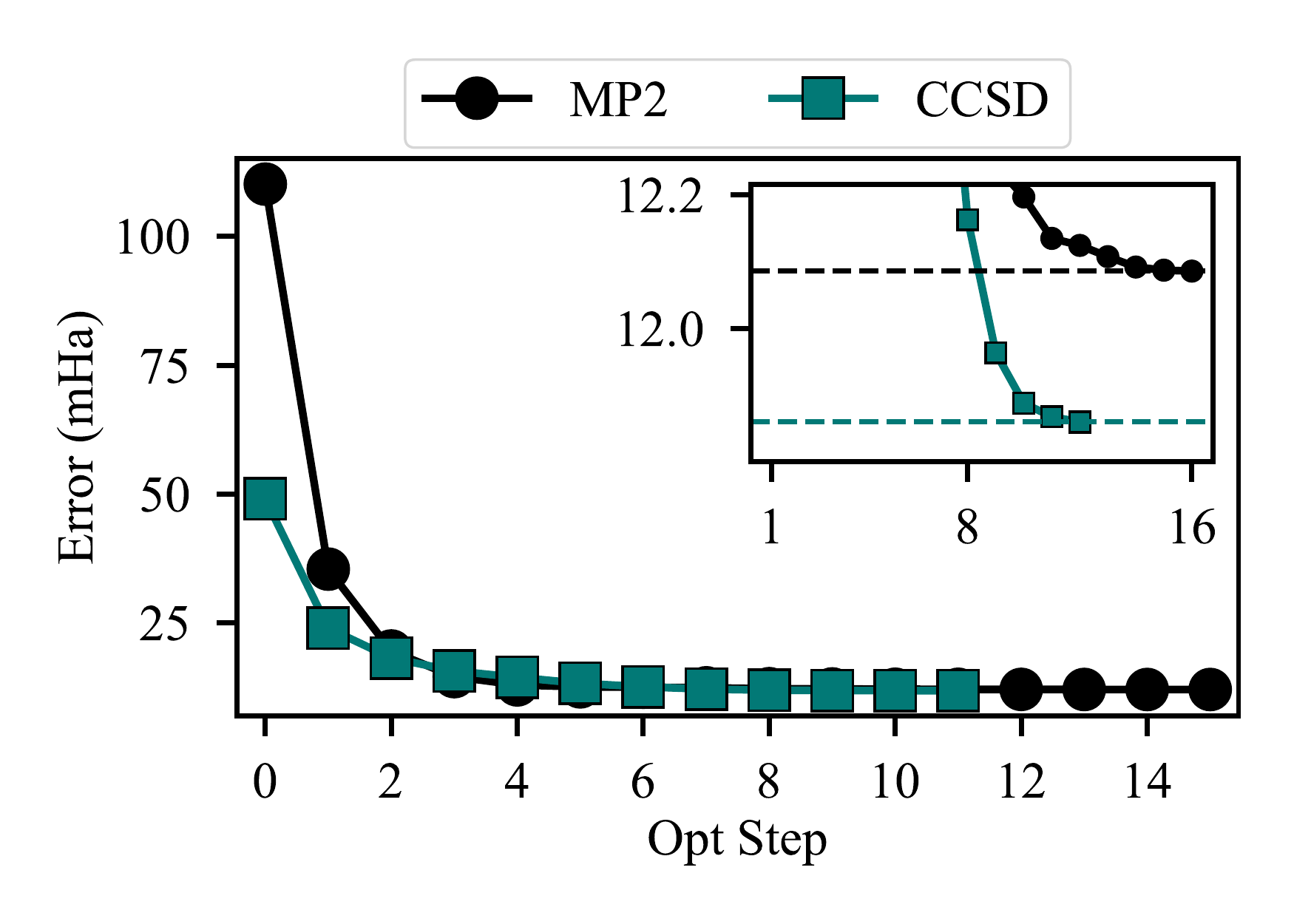}}
    \\
    \subfloat[][Stretched LiH]{\includegraphics[width=0.3\linewidth]{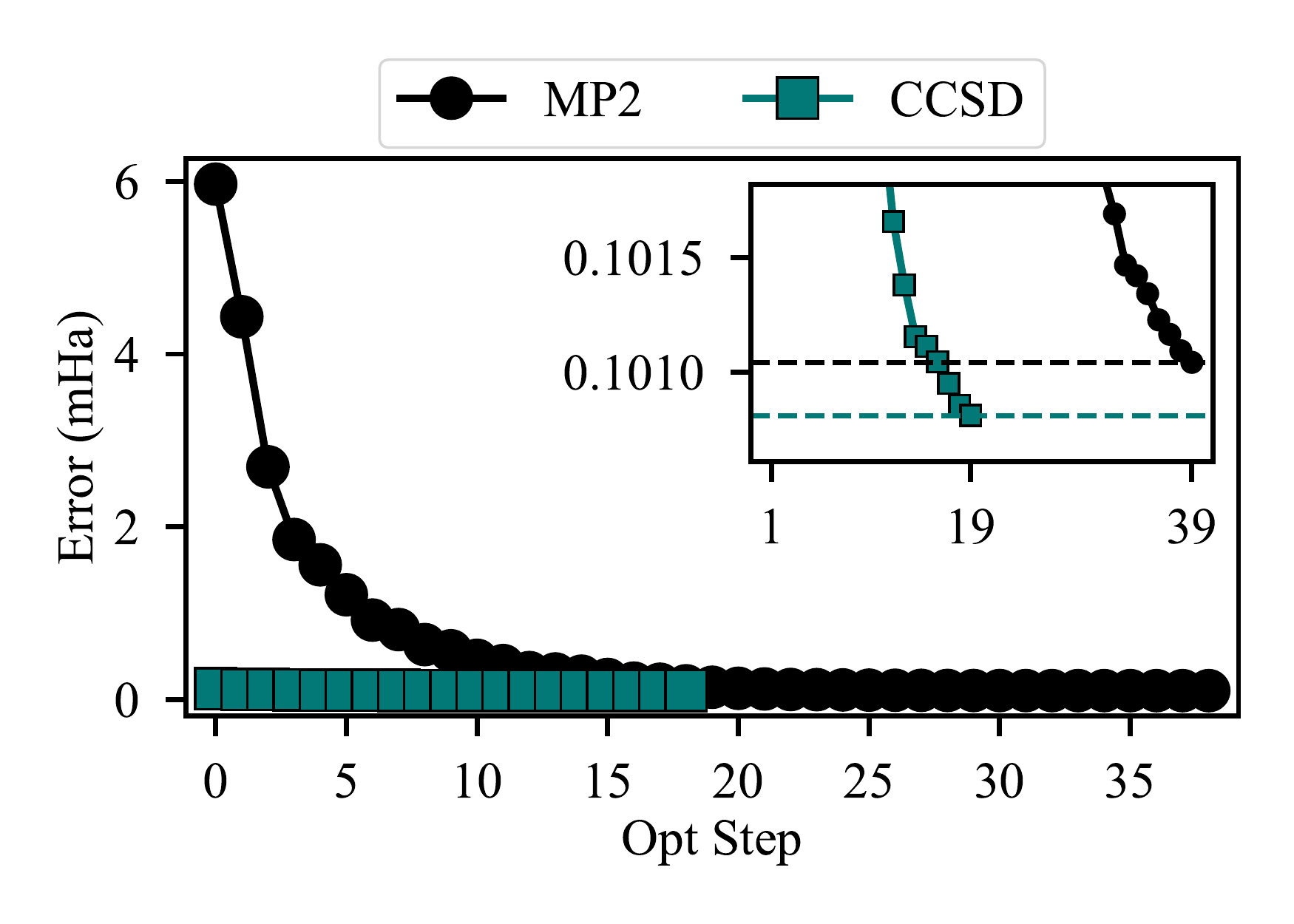}}
    \subfloat[][Stretched $\mathrm{BeH}_2$]{\includegraphics[width=0.3\linewidth]{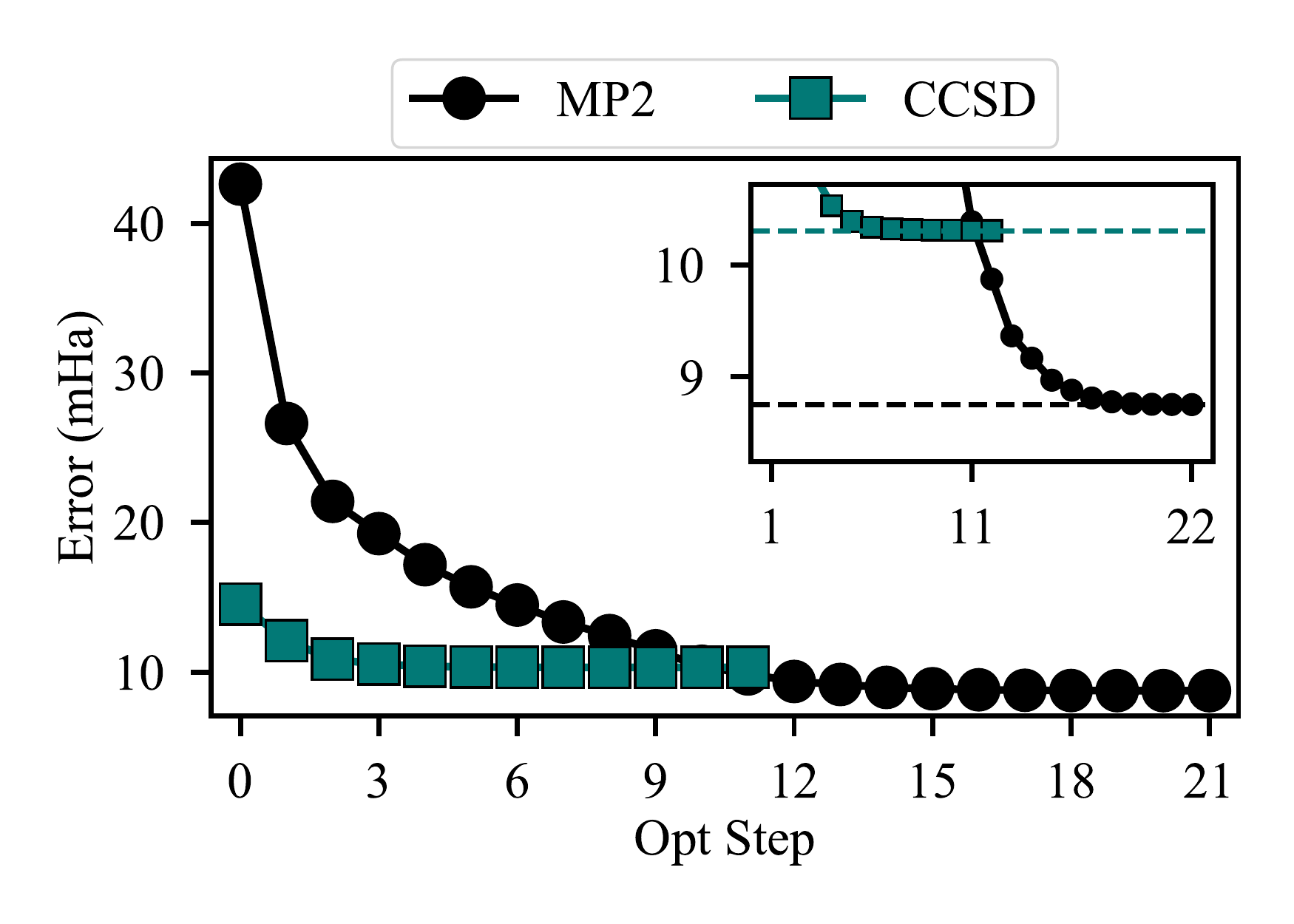}}
    \subfloat[][HF]{\includegraphics[width=0.3\linewidth]{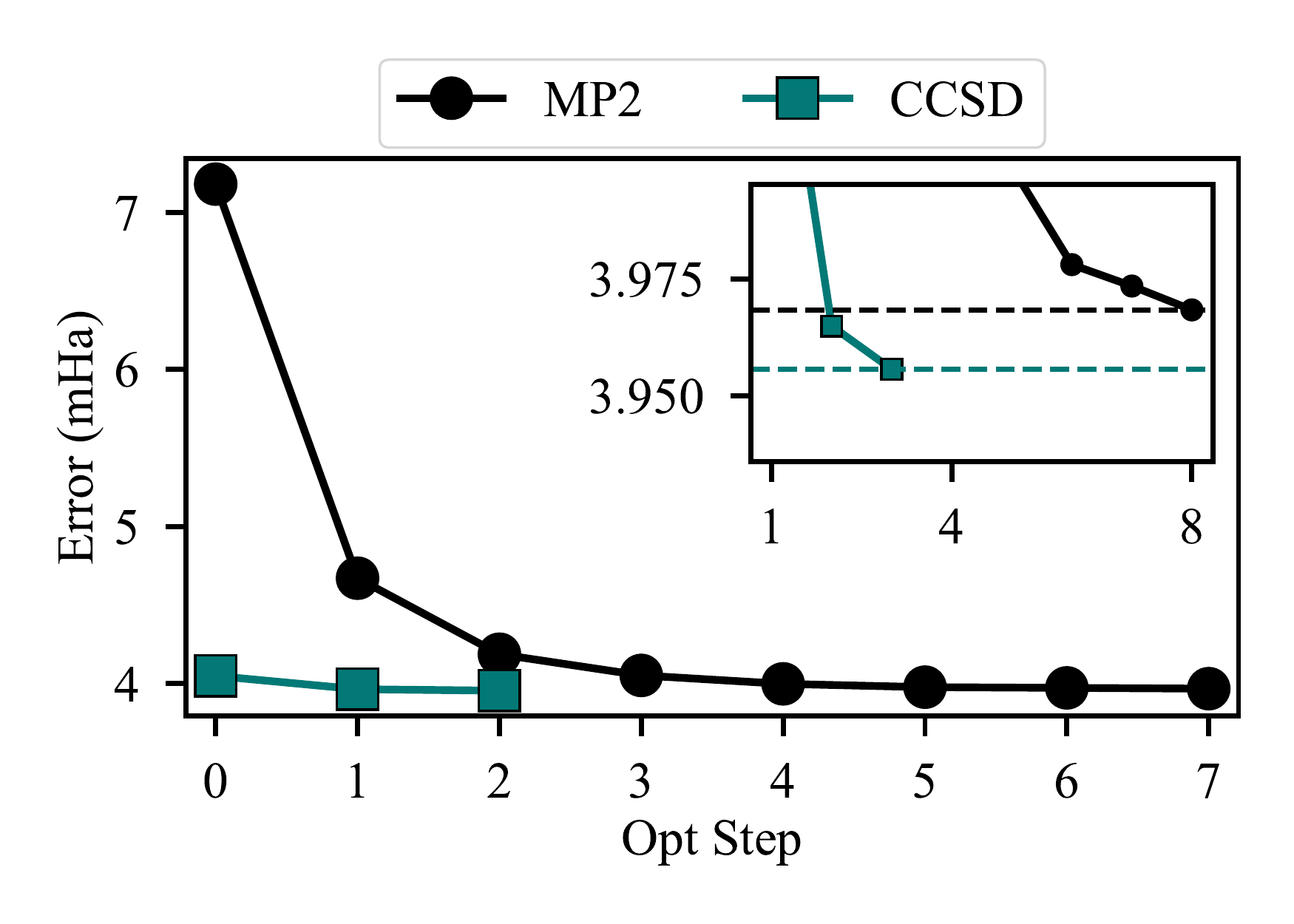}}
    \caption[]{The convergence of the groundstate energy for equilibrium and stretched LiH, equilibrium and stretched $\mathrm{BeH}_2$, stretched $\mathrm{H}_{10}$, and HF starting from circuits initialized with MP2 (black circles) and CCSD (green squares) parameters. The insets in show the same data on a smaller scale to better resolve the errors after many optimization steps. The errors are calculated relative to the FCI energy except for HF, in which it is relative to the ASCI energy.}
    \label{fig:convergence}
\end{figure*}

Finally, we briefly compare classical optimization of the UCC ansatz initialized with MP2 and CCSD parameters. As test cases we study LiH, $\mathrm{BeH}_2$, HF, and stretched $\mathrm{H}_{10}$, again using experimental geometries~\cite{CCCBDB}. We consider both equilibrium and stretched geometries for LiH and $\mathrm{BeH}_2$, stretched by factors of 1.5 and 2.0 respectively, and the same stretched geometry of $\mathrm{H}_{10}$ as above. We use the cc-pCVDZ basis for LiH, the cc-pVDZ basis and the frozen core approximation for $\mathrm{BeH}_2$ and HF, and the STO-6G basis for stretched $\mathrm{H}_{10}$. We note that although the MP2 initialization sets $\theta=0$ for all singles operators, the circuits still contain these operators and their angles are included in the optimization. For each calculation we use $N_{\mathrm{WF}}=100,000$ and do not attempt to extrapolate energies to the infinite-$N_{\text{WF}}$ limit. Since the molecules in this study all have singlet spin states and geometries with symmetry, we reduce the number of variational parameters that need to be optimized in the UCCSD ansatz in two ways. First, we use spin-complemented single- and double-excitation operators that combine parameters that are necessarily the same for these electronic states~\cite{Grimsley:2019:3007}. Second, we use point group symmetry arguments to eliminate parameters associated with operators that do not retain the symmetry of the initial Hartree-Fock SCF solution~\cite{Cao:2022:062452}. This reduces the number of parameters by a factor of approximately the order of the point group. For example, these symmetry considerations reduce the number of parameters in the LiH circuit from 2268 to 408. We use the BFGS method as implemented in SciPy to optimize the circuit parameters, setting the tolerance to $10^{-3}$, except for HF for which we set it to $5\times10^{-3}$ due to limited computational resources~\cite{scipy}. We plot the errors as a function of optimization step in Fig.~\ref{fig:convergence}.

For the weakly correlated systems (equilibrium LiH, equilibrium $\mathrm{BeH}_2$, and stretched LiH), we find that the optimizer fails to significantly reduce the error produced by the CCSD-initialized circuit, indicating that these parameters are near optimal. The remaining, more strongly correlated systems exhibit significant reduction of the error with optimization of the CCSD-initialized circuits. The MP2- and CCSD-initialized circuits obtain approximately the same final error after optimization, but the MP2 circuits consistently require many more optimization steps to achieve this minimum error. Considering current difficulty of classical optimization of VQE circuits implemented on NISQ-era hardware, we posit that initializing the UCC ansatz with CCSD parameters rather than MP2 parameters can significantly reduce the challenge of achieving quantum advantage in VQE. We also note that the final error obtained by the MP2 circuit for stretched $\mathrm{BeH}_2$ is significantly lower than the CCSD circuit. This indicates that the MP2 circuit lays in a basin of attraction around a better local minimum than the CCSD circuit, despite having a larger initial error.  This result indicates the relationship between initial energy and the nearest local minima of the energy landscape for the UCC ansatz is deserving of further study. Regardless, even in this case, the CCSD circuit reaches the minimum of its basin of attraction in significantly fewer optimization steps than the MP2 circuit takes to reach its own basin.

\textit{Discussion}.--- In this paper we demonstrated through extensive calculations that CCSD parameterizations of the UCC ansatz consistently outperform their MP2 counterparts.  As such, it is important to compare the computational costs of obtaining the CCSD and MP2 parameterizations. Although MP2 is faster and, in fact, often used as a starting point for coupled cluster simulations, CCSD nevertheless requires reasonable classical computation resources for even moderately sized systems. For example, the CCSD calculations presented in this work and others run in minutes or less on a laptop~\cite{Baek2022,Fed2021}.

MP2 and CCSD runtimes scale as O($N^5$) and O($N^6$), respectively, making these prohibitively expensive algorithms in the large-$N$ qubit limit, but it is unlikely that NISQ era quantum computers will exceed classically-accessible simulations of CCSD in the near future.  Classical coupled cluster simulations can be accelerated in various ways~\cite{rip2013,kaliman2017}, indicating that simulations involving hundreds of qubits to parameterize circuits is in reach. Considering this, as well as the small prefactors of these runtime scalings and the efficiency of modern implementations of these techniques, CCSD is poised to remain an accessible and highly accurate method of UCC parameterization for the forseeable future of the NISQ era. As such, our results suggest that CCSD should replace MP2 as the standard approach to classically parameterizing UCC circuits. 

In our investigation we found that circuits initialized with CCSD parameters produce drastically lower energies than those initialized with MP2 parameters. In fact, for weakly correlated molecules the CCSD-initialized circuits often obtain nearly all of the correlation energy. Furthermore, we showed that the energies produced by circuits for strongly correlated molecules converge in far fewer optimization steps when initialized with CCSD rather than MP2 parameters. Despite differences in convergence rate, we found in all cases but one that the MP2- and CCSD-initialized circuits obtain similar energies after optimizations. In one case, BeH$_{2}$, the MP2-initialized circuit obtained a significantly lower energy after optimization than the CCSD-initialized circuit, indicating that the two initial circuits lie in the different basins of attraction, an interesting result about local minima deserving of further investigation.

Our results also display the power of our sparse wavefunction circuit solver, which enables us to perform UCC simulations at system sizes that have not been previously explored. Because our solver is capable of handling up to 64 qubit problems with its current implementation, we are able to access a regime in which it is possible to meaningfully test and differentiate VQE results. In this case, the ability to access large systems sizes enabled us to explore a widely used parameterization for UCC circuits and challenge conventional knowledge about it.

Testing our approach with higher order coupled cluster techniques on both the classical~\cite{tubman2020} and quantum side~\cite{Fed2021} is an important topic of future research. The correspondence we identified between CCSD and UCC (CCSD) is weakened when classical CCSD breaks down, as seen in for strongly correlated molecules like stretched $\mathrm{H}_{10}$. These results motivate the study of more advanced classical approaches to parameterize UCC-type circuits. Establishing the correspondence between higher order classical coupled cluster theories and the UCC analogues of them, such as a UCC (CCSDT) circuit~\cite{Fed2021}, would elucidate the full potential of the UCC ansatz.

\textit{Acknowledgements}.--- We thank Diptarka Hait for assisting our  calculations. We are grateful for support from NASA Ames Research Center. We acknowledge funding from the NASA ARMD Transformational Tools and Technology (TTT) Project. Part of this work is funded by U.S. Department of Energy, Office of Science, National Quantum Information Science Research Centers, Co-Design Center for Quantum Advantage under Contract No. DE-SC0012704. Calculations were performed as part of the XSEDE computational Project No. TG-MCA93S030 on  Bridges-2 at the Pittsburgh supercomputer center. M.R.H. and D.C. were supported by NASA Academic Mission Services, Contract No. NNA16BD14C.  M.R.H. and D.C. participated in the Feynman Quantum Academy internship program.


%

\end{document}